\documentclass[traditabstract]{aa} 
\usepackage{epsfig}
\usepackage{setspace}
\usepackage{graphicx}
\usepackage{xcolor}
\usepackage{bm}
\usepackage{amsmath}
\usepackage{txfonts}
\usepackage{amssymb}
\usepackage{natbib}                          	                

\def\apjl{ApJL }

\def\apjs{ApJS }

\def\aap{A\&A }

\def\be{\begin{equation}} 
\def\ee{\end{equation}} 

\begin{document} 

\title{Approximate analytical calculations of photon geodesics in the Schwarzschild metric}
 
\titlerunning{Approximate Schwarzschild photon geodesics}  
\authorrunning{V. De Falco}  
  
\author{Vittorio De Falco$^{1,2}$\and  
Maurizio Falanga$^{2,1,3}$\and 
Luigi Stella$^{4}$}

\institute{Institut f\"ur theoretische Physik, Universit\"at Basel, Klingelbergstrasse 82, 4056 Basel, Switzerland\\
\email{vittorio-df@issibern.ch} 
\and International Space Science Institute, Hallerstrasse 6, 3012 Bern, Switzerland 
\and International Space Science Institute Beijing, No.1 Nanertiao, Zhongguancun, Haidian District, 100190 Beijing, China
\and INAF -  Osservatorio Astronomico di Roma,  Via Frascati, 33, Monteporzio Catone, Rome, 00078, Italy
} 
 
\date{} 

\abstract{We develop a method for deriving approximate analytical formulae to integrate photon geodesics in a Schwarzschild spacetime. Based on this, we derive the approximate equations for light bending and 
propagation delay that have been introduced empirically. We then derive for the first time an approximate analytical equation for the solid angle. We discuss the accuracy and range of applicability of the new equations and present a few simple applications of them to known astrophysical problems.}


\maketitle 
\section{Introduction}
\label{sec:intro}
Early studies \citep{Luminet79, Pechenick83} began a great interest in photons emitted by matter in a strong gravitational field, especially in relation to high-energy astrophysics. The relevant computations are carried out with ray-tracing techniques that are based on the photon geodesics in general relativistic spacetimes. Effects to be considered are ({\it i}) light bending, ({\it ii}) travel time delay, and ({\it iii}) gravitational lensing \citep{Misner73}. The basic equations for the Schwarzschild metric are expressed through elliptic integrals that can be solved numerically. A powerful analytical approximation was introduced by \citet{Beloborodov02}, who derived an approximate linear equation to describe the gravitational light bending of photons emitted at radius $r \geqslant r_s$ ($r_s = 2GM/c^2$).  
In the same vein, \citet{Poutanen06} derived an approximate polynomial equation for photon travel time delays. These two analytical approximations were obtained by introducing an {\it \textup{ad hoc}} parametrization of the photon emission angle \citep[see][for more details]{Beloborodov02,Poutanen06}. Nevertheless, the equation for gravitational lensing, also known as solid angle equation, was still solved numerically by these authors. 

In this paper we present a mathematical method through which the approximate polynomial equations for light bending and travel time delay in a Schwarzschild spacetime are derived without any ad hoc assumption. We then apply the same method to derive for the first time an approximate polynomial equation for gravitational lensing. High-accuracy approximate equations for photon geodesics translate into high-speed ray-tracing codes for different astrophysical applications in the strong gravitational field of Schwarzschild black holes (BHs). As examples we apply our approximate equations to calculate the light curve from a hot spot on the surface of a rotating neutron star (NS) and a clump in a circular orbit around  BH. Moreover, we calculate the fluorescent iron $K{\alpha}$ line profile from an accretion disk around a BH \citep[e.g.,][]{Fabian89}.

\section{Photons in the Schwarzschild spacetime}
\label{sec:equations}
In this section we introduce the elliptical integrals of photon trajectories, travel time delay, and gravitational lensing in the Schwarzschild metric. 
\begin{figure}[h]
\centerline{\psfig{figure=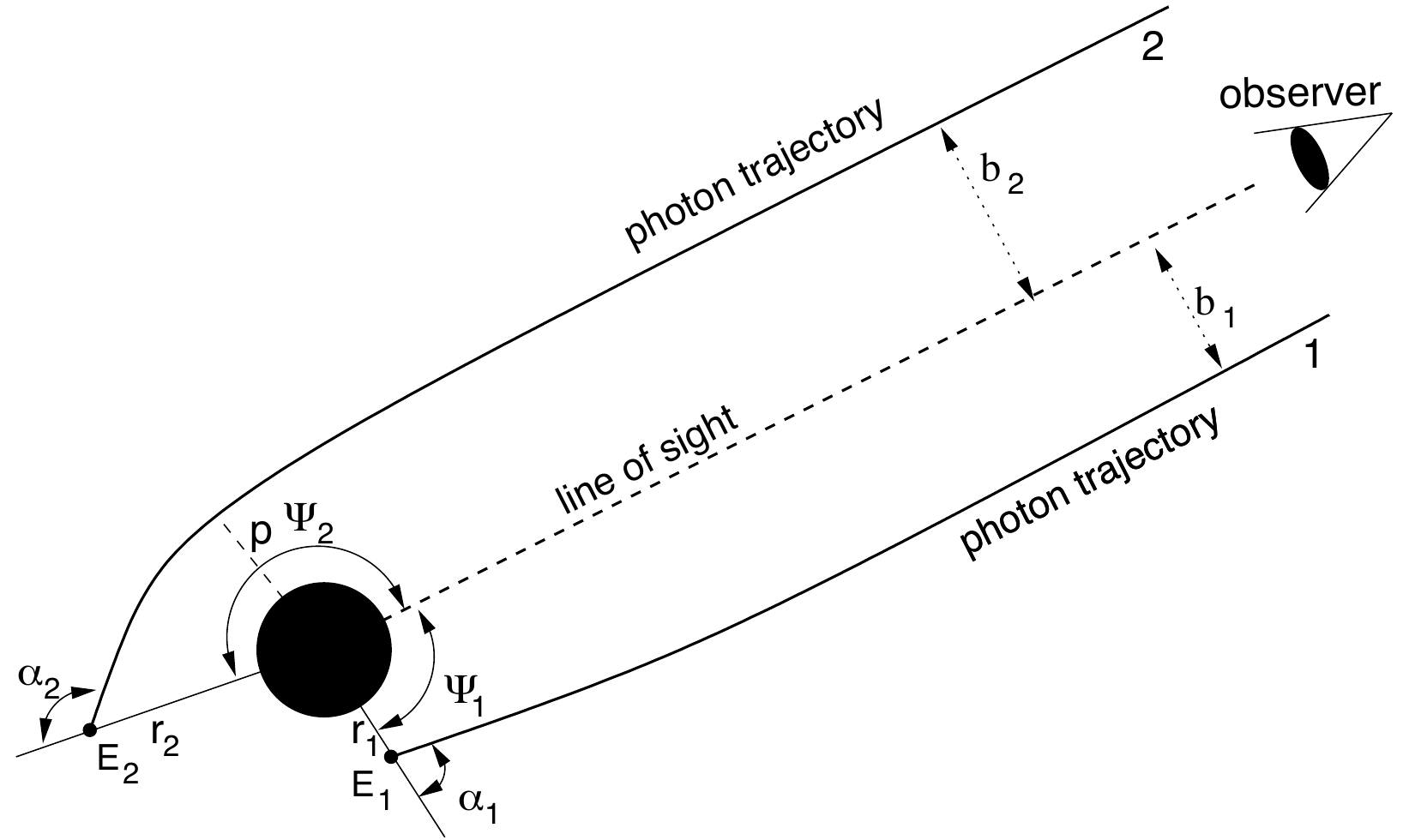, width=7.5cm}} 
\caption{Two photon trajectories emitted at different radii, $r_1$ and $r_2$, and emission angles, 
$\alpha_1$ and $\alpha_2$ , with their corresponding impact parameters, $b_1$ and $b_2$. Trajectory 1 is for a direct photon, while trajectory 2 has a turning point, i.e., passes through periastron $p$, the minimum distance between the trajectory and the BH. The observer is at infinity, and photons geodesics lie in a single invariant plane.} 
\label{fig:Fig1} 
\end{figure}

\subsection{Schwarzschild metric}
\label{subsec:Schwarzschild}
For static and spherically symmetric BHs of mass, $M$, the Schwarzschild metric in spherical coordinates ($t, r, \varphi, \psi$) is
\begin{equation} 
ds^2=\left (1-\frac{r_{\rm s}}{r} \right ) dt^2 - \left (1-\frac{r_{\rm s}}{r} \right )^{-1}dr^2-r^2(d\varphi^2+ \sin^2\varphi\, d\psi^2),
\label{eq:sm} 
\end{equation}
where $G=c=1$, and $r_{\rm s}=2M$ is the Schwarzschild radius. In this standard system, the coordinate variables are time $t$, radius $r$, polar angle $\varphi$, and azimuthal angle
$\psi$. 

\subsection{Gravitational light bending}
\label{subsec:impa&libe}
Because of spherical symmetry, it is customary to use the equatorial plane at $\varphi=\pi/2$ to calculate geodesics in the Schwarzschild metric that are representative of all photon trajectories. A photon geodesic starting at radius $R$ is described by the following elliptical integral \citep{Chandrasekhar92,Misner73}: 
\begin{equation} \label{eq:libe} 
\psi=\int_R^\infty \frac{dr}{r^2}\left[\frac{1}{b^2}-\frac{1}{r^2}\left(1-\frac{r_s}{r} \right) \right]^{-\frac{1}{2}},
\end{equation}
parametrized by the ratio of the angular momentum, $L$, and energy, $E$, of the photon, $b=L/E$.  
The  impact parameter $b$ represents the distance between the observer and the photon trajectory at infinity and is related to the photon emission angle, $\alpha$ by
\begin{equation} \label{eq:impact} 
b = \frac{R\sin\alpha}{\sqrt{1-r_{s}/R}}.
\end{equation}
Equation (\ref{eq:libe}) is strictly valid up to $\alpha=\pi/2$, since the sine function is symmetric with respect to $\alpha=\pi/2$. The photon deflection angle, $\psi$, can be directly determined in terms of the emission angle $\alpha$ through Eq. (\ref{eq:impact}). 

We must distinguish between direct photons, which have trajectories with an emission angle between $0\le\alpha\le\pi/2$, and photons with a turning point, whose trajectories have an emission angle ranging between $\pi/2\le\alpha\le\alpha_{max}$ (see Fig. \ref{fig:Fig1}). Photon trajectories with a turning point can reach infinity only if their $b$ is greater than the critical impact parameter $b_c=3\sqrt{3}M$ \citep[see, e.g.,][]{Luminet79}. Since we are interested only in photons that are not captured by the BH, the maximum possible emission angle is obtained by substituting $b_c$ into Eq. (\ref{eq:impact}) 
\begin{equation} \label{alphamax}
\alpha_{\rm max}=\pi-\arcsin\left[\frac{3}{2}\sqrt{3\left(1-\frac{r_{s}}{R}\right)}\frac{r_{s}}{R}\right].
\end{equation}   
Photons emitted between $\pi/2\le\alpha\le\alpha_{\rm max}$ follow trajectories with a turning point; therefore a periastron distance, $p$, is defined at an angle $\alpha_{\rm p}=\pi/2$, which determines the minimum distance between the compact object and the photon trajectory. The emission point of a photon at $\psi_{\rm E}$ that passes through the turning point is symmetric with respect to the periastron angle, $\psi_{\rm p}$, to the point $\psi_{\rm S}$, (with an emission angle $\alpha\le\pi/2$) along the same trajectory, as they have the same impact parameter at infinity. Based on this symmetry, we determine  $\psi_{\rm S} = 2\psi_{\rm p}-\psi_{\rm E}$, where $\alpha_{\rm S}=\pi-\alpha_{\rm E}$. 

\subsection{Travel time delay}
\label{subsec:tide}
A photon following its geodesic from an emission point, $E$, to an observer at infinity has an infinite travel time, $\Delta \tau$, value. To have a finite quantity, we calculate the relative travel time delay between a photon emitted at a distance, $R$, following its geodesic and the photon emitted radially with $b=0$, that is, $\Delta t(b)=\Delta \tau(b) - \Delta \tau(b=0)$ \citep{Pechenick83}. In the Schwarzschild metric we have
\begin{equation} \label{eq:tide} 
\Delta t =\int_R^\infty \frac{dr}{1-\frac{r_s}{r}} \left \{ \left [ 1-\frac{b^2}{r^2}\left(1-\frac{r_s}{r} \right) \right ]^{-\frac{1}{2}}  -1  \right \}.
\end{equation}

\begin{figure}[ht]
\centerline{\psfig{figure=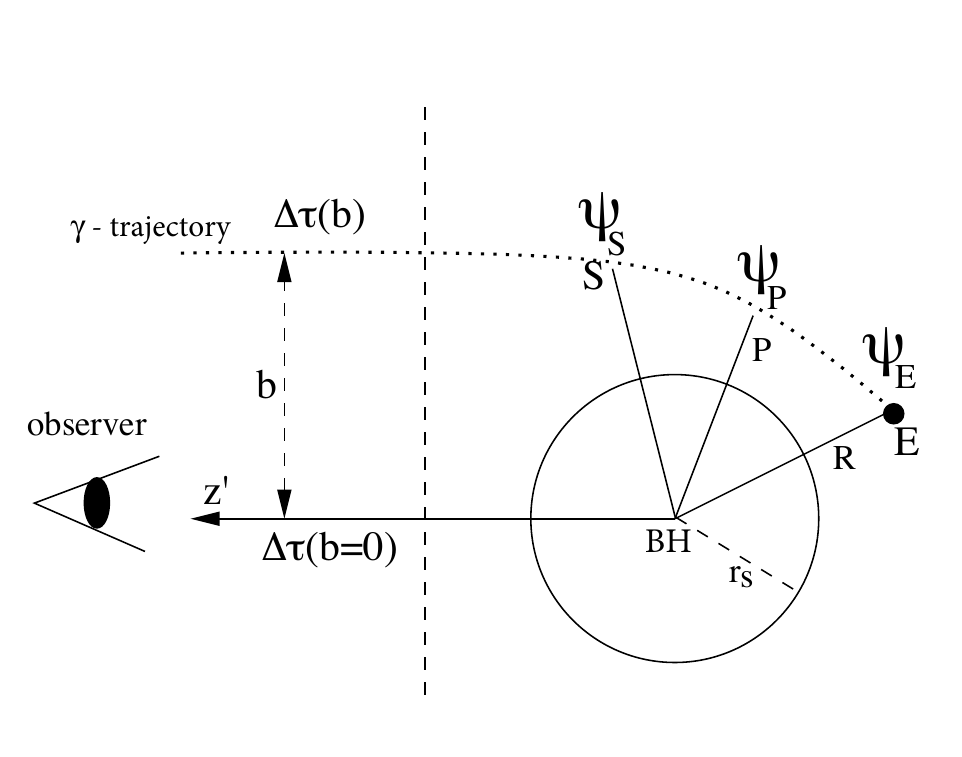, width=5.5cm}} 
\caption{Calculation of travel time delay for trajectories with turning points. The photon is emitted at $E$ with radius $R$ and deflection angle $\psi_{\rm E}$. The photon trajectory passes through point, $S$, which is symmetric to $E$ with respect to periastron $p$, having the same impact parameter, $b$, and a deflection angle $\psi_{\rm S} = 2\psi_{\rm p}-\psi_{\rm E}$.} 
\label{fig:Fig1_1}
\end{figure}

To calculate the time delay for photons with a turning point, we need to calculate the periastron distance, $p$. For a given $b$ we therefore consider the largest real solution of the following equation $p^3-b^2p+b^2r_s=0$. The polynomial in $p$ has three real solutions (because $b \ge b_{\rm c}$):  one is negative, one is lower than $3M,$ and we consider only the solution satisfying $p\ge r_c$, where $r_c=3M$ is the critical radius associated to $b_c$ \citep[see, e.g.,][]{Luminet79}. The time delay is composed of the time delay $\Delta t_{\rm S}$ from point $\alpha_{\rm S}$, as determined by the Eq. (\ref{eq:tide}), plus the time delay between $[\alpha_{\rm S}, \alpha_{\rm p}]$, $\Delta t_{\rm p-S}$, and $[\alpha_{\rm p}, \alpha_{\rm E}]$, $\Delta t_{\rm E-p}$. Since the integrand is symmetric with respect to $\alpha_p$ , the latter two time delays are equal ($\Delta t_{\rm E-p} = \Delta t_{\rm p-S}$), the equation can be written (see Fig. (\ref{fig:Fig1_1}) ) 
\begin{equation} \label{eq:tidetp} 
\begin{aligned}
\Delta t &=\Delta t_{\rm S}+2\Delta t_{\rm p-S} =\\
&=\Delta t_{\rm S}+2\int_R^p \frac{dr}{1-\frac{r_s}{r}} \left \{ \left [ 1-\frac{b^2}{r^2}\left(1-\frac{r_s}{r} \right) \right ]^{-\frac{1}{2}}\right \}.
\end{aligned}
\end{equation}

\subsection{Solid angle}
\label{subsec:soan}
We consider the emission reference frame of coordinates $(x,y,z)$ and the observer reference frame of coordinates $(x',y',z')$, where the two systems are rotated with an angle, $i$, around $y=y'$. The solid angle, $d\Omega$, in the observer reference frame reads as $d\Omega=\sin\psi \,d\psi \, d\varphi$. This equation can be expressed in terms of the impact parameter, $b$, by its first-order approximation for infinitesimally small $\psi$ as $b\approx D\cdot \psi$, where $D$ is the distance from the emission point to the observer,
\begin{equation} 
\label{CASF} 
d\Omega=\frac{b\ db\ d\varphi'}{D^2}.
\end{equation}
In the emission reference frame, Eq. (\ref{CASF}) becomes
\begin{equation} \label{NCASF} 
d\Omega=\frac{b}{D^2}\frac{\partial \varphi'}{\partial \varphi}\frac{\partial b}{\partial r} dr d\varphi,
\end{equation}
where we considered the following dependencies $\varphi=\varphi(\varphi')$ and $b=b(r,\psi)$. The Jacobian of the transformation is always $\frac{\partial \varphi'}{\partial \varphi}\frac{\partial b}{\partial r}$ independent of the value of $\frac{\partial b}{\partial \psi}$, since the photon moves in an invariant plane. Therefore, Eq. (\ref{NCASF}) is valid for any emission point. To calculate the Jacobian, we use the following coordinates transformation $\cos\psi=\sin i\cos\varphi$ that relates the angles in the observer and emission reference frames. $\frac{\partial b}{\partial r}=-\frac{\partial b}{\partial \psi}\frac{\partial \psi}{\partial r}$ is calculated using the light bending Eq.  (\ref{eq:libe}). The solid angle equation in the Schwarzschild metric is thus \citep[see, e.g.,][]{Bao1994}\footnote{Equation (\ref{EFSA}) is equivalent to the formula (A3) in \citet{Beloborodov02}.}

\begin{equation} \label{EFSA} 
d\Omega=\frac{\frac{\cos i}{D^2\ R^2\ \sin^2\psi}\frac{b^2}{\cos\alpha}} {\int_R^\infty \frac{dr}{r^2}\left[1-\frac{b^2}{r^2}\left(1-\frac{r_s}{r} \right) \right]^{-\frac{3}{2}}}\ dr\ d\varphi.
\end{equation}
This equation contains an integral with the same functional form as those of light bending Eq. (\ref{eq:libe}) and time delay Eq. (\ref{eq:tide}), except for the $-3/2$ exponent and factors depending on the impact parameter $b$ (or emission angle $\alpha$).   

\section{Analytical approximations}
\label{sec:Method}
In this section we present the general mathematical method used to approximate the elliptical equations in polynomials of light bending Eq. (\ref{eq:libe}), time delay Eq. (\ref{eq:tide}), and solid angle Eq. (\ref{EFSA}). 

\subsection{Mathematical method}
\label{subsec:matmet}
Let $f$ be an integrable function of radius, $r$, mass, $M$, and sine of the emission angle, $\sin\alpha$, that is, $f=f(r,M,\sin\alpha)$ and $I$ the following elliptic integral
\begin{equation} \label{INT}
I=\int_{r_i}^{r_f} \frac{1}{\sqrt{f(r,M,\sin\alpha)}}\ dr.
\end{equation} 
We are interested in deriving a polynomial approximation of the elliptic integral $I$. We first define $\sin\alpha=g(z)$, where $g(z)$ is a generic function of $z(\alpha)$. To expand Eq. (\ref{INT}) in Taylor series we assume that $\alpha$ is very small\footnote{Therefore, $g(z)$ is small as well.}  and aim at obtaining an integrable polynomial function
\begin{equation} \label{POL} 
I=\int_{r_i}^{r_f} \frac{1}{\sqrt{f(r,M,g(z))}}\ dr\approx P(r_f,r_i,M,g(z)).
\end{equation} 
$P$ contains even powers of $g(z)$, since $f(r,M,g(z))\propto g(z)^2$. This condition is given by substituting $b = (r\sin\alpha) / (\sqrt{1-r_{s}/r})$ in the equations of the light bending Eq. (\ref{eq:libe}), time delay Eq. (\ref{eq:tide}), and solid angle Eq. (\ref{EFSA}). For an exact polynomial approximation, we therefore define $g(z)=\sqrt{Az^2+Bz}$, where $A$ and $B$ are general parameters. One of the two parameters ($A, B$) is determined by comparing Eq. (\ref{POL}) with the original integral $I$ for special values of $M=M*$, $r_f=r_f*$ and $r_i=r_i*$ that permits solving the integral $I$ easily and obtain

\begin{equation} \label{EQL} 
I(r_f*,r_i*,M*,\sin\alpha)=P(r_f*,r_i*,M*,\sqrt{Az^2+Bz}).
\end{equation}  
The other parameter can be determined through the initial condition $\sin\alpha=\sqrt{Az^2+Bz}$. We note that the polynomial approximation is valid for any emission angle $\alpha$ (not only for low values) since the parameters $A, B$ are gauged on the whole range of $I$.

\subsection{Light bending}
\label{sec:applibe}
For the light bending we Taylor-expand Eq. (\ref{eq:libe}) up to the third order and defining $u=2M/R$ and $\sin\alpha=g(z)$ we obtain
\begin{equation}\label{eq:lb}
\begin{aligned}
\psi\approx \frac{b}{R}&\left[1+\frac{g^2(z)}{6(1-u)}-\frac{g^2(z)u}{8(1-u)}+\frac{3g^4(z)}{40(1-u)^2}+\right.\\
&+\frac{3g^4(z)u^2}{56(1-u)^2}-\frac{g^4(z)}{8(1-u)^2}u+\frac{5g^6(z)}{112(1-u)^3}-\\
&\left. -\frac{g^6(z)u^3}{32(1-u)^3}-\frac{15g^6(z)}{128(1-u)^3}u+\frac{5g^6(z)}{48(1-u)^3}u^2\right].
\end{aligned}
\end{equation}
Setting $g(z)=\sqrt{Az^2+Bz}$ and neglecting all the terms up to the second order in $z$, Eq. (\ref{eq:lb})  
becomes
\begin{equation} \label{TEF} 
\psi\approx\sqrt{\frac{Az^2+Bz}{1-u}}\left[1+\left(\frac{B}{6(1-u)}-\frac{Bu}{8(1-u)}\right)z \right].
\end{equation}
To approximate this equation with a polynomial, we introduce an even trigonometric function of $\psi$ to remove the square root. The simplest choice is a cosine function expanded to the fourth order in $\psi$

\begin{equation} \label{COSESP} 
\begin{aligned}
1-\cos\psi&\approx \frac{\psi^2}{2}-\frac{\psi^4}{24}\approx\\
&\approx \frac{Bz}{2(1-u)}+\left [\frac{B^2}{6(1-u)^2}-\frac{B^2u}{8(1-u)^2}+\right .\\
&\quad\left. +\frac{A}{2(1-u)}-\frac{B^2}{24(1-u)^2} \right ]z^2,
\end{aligned}
\end{equation}
where we consider the terms to the second order in $z$. If we choose $A=-(B/2)^2$ , we obtain a simple linear approximation, $1-\cos\psi \,\approx Bz/(2(1-u))$, in which $z^2$ coefficients vanish. 

We now solve Eq. (\ref{eq:libe}) for the special values $u=0,\ R=1$ and obtain
\begin{equation} \label{TELB} 
\psi=b\int_1^\infty \frac{dr}{r^2}\left[1-\frac{\sin^2\alpha}{r^2} \right]^{-\frac{1}{2}}=\alpha.
\end{equation}
Using the same values ($u=0,\ R=1$) for the approximated polynomial equation, $1-\cos\psi \,\approx Bz/(2(1-u))$, we obtain
\begin{equation}
1-\cos\alpha=\frac{Bz}{2}.
\end{equation}
In this case, by defining $B=2$ (implying $A=-1$), we find $z=1-\cos\alpha$, which, when replaced in Eq. (\ref{COSESP}), gives the approximate light bending equation originally found by \cite{Beloborodov02}
\begin{equation} \label{AFLB} 
1-\cos\psi=\frac{(1-\cos\alpha)}{(1-u)}.
\end{equation}

\begin{figure}[h]
\centering
\centerline{\psfig{figure=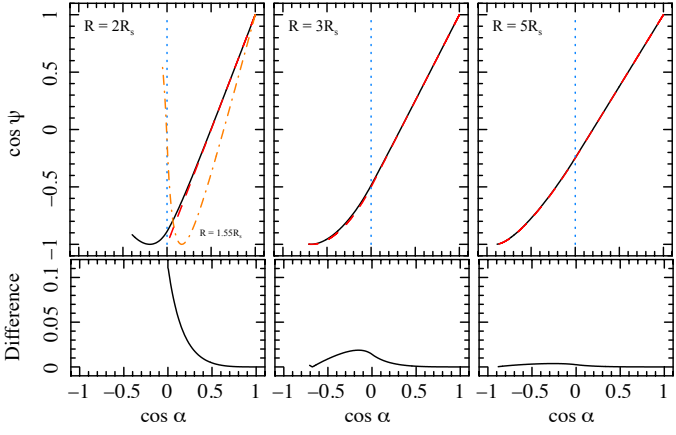,trim=0.5 3cm 0cm 3cm,scale=0.38}} 
\caption{Light bending curves from the exact Eq. (\ref{eq:libe}) (solid lines), compared to those from the approximate Eq. (\ref{AFLB}) (dashed red lines) for $R=2r_s$, $R=3r_s$, and $R=5r_s$. The dotted blue line represents the threshold from trajectories without a turning point ($0\le\alpha\le\pi/2$) to trajectories with a turning point ($\pi/2\le\alpha\le\alpha_{max}$). The exact light bending curve for $R=1.55r_s$ is also plotted (dotted-dashed orange line) to show strong-field effects. The lower panels show the difference between the curves from the original and approximate equations.} 
\label{fig:Fig2}
\end{figure}

In Fig. (\ref{fig:Fig2}) we show a comparison between the exact light bending curves for different emission radii, and curves obtained from the approximate equation. The accuracy of the latter between $0\le\alpha\le\alpha_{max}$ is better than 3\% for $R=3r_s$, while for $R = 5r_s$ the error does not exceed 1\%. We note that $R = 3r_s$ corresponds to the innermost stable circular orbit (ISCO) for matter orbiting a Schwarzschild BH and is also close represent to a typical NS radius size of $\sim12$ km for mass of $1.4M_{\odot}$. For values below $R=2 r_s$ the equation is not anymore applicable after $\alpha = \pi/2$. In Fig. \ref{fig:Fig2} we also show the exact light bending curve for $R=1.55r_s$; after a given minimum the photons are highly bent by strong-field effects. The largest error is at $\alpha=\pi/2$ and then it tends to decrease until at $\alpha_{max}$ because of the symmetrization process around $\alpha=\pi/2$ configuring as the maximum reachable angle (see Sect. \ref{subsec:impa&libe}). For more details about the accuracy between $0\le \alpha \le \pi/2$ we refer to \citet{Beloborodov02}.

\subsection{Time delay}
\label{sec:apptide}
We now apply our method for deriving the approximate equation for the time delay. By expanding the integrand in Eq. (\ref{eq:tide}) up to the third order 
\begin{equation} \label{TDGE2}
\begin{aligned}
\Delta t = R&\left \{\frac{g^2(z)}{2(1-u)}+\frac{g^4(z)}{8(1-u)^2}-\frac{3g^4(z)}{32(1-u)^2}u+\right.\\
&\left. +\frac{g^6(z)}{16(1-u)^3}-\frac{5g^6(z)}{48(1-u)^3}u+\frac{5g^6(z)}{112(1-u)^3}u^2\right \},
\end{aligned}
\end{equation}   
we set again $g(z)=\sqrt{Az^2+Bz}$ and neglect all terms up to the third order in $z$, so that
\begin{equation} \label{TDGE3}
\begin{aligned}
\frac{\Delta t}{R} \approx &\left \{\frac{Az^2+Bz}{2(1-u)}+\frac{B^2z^2+2ABz^3}{8(1-u)^2}-\frac{3u(B^2z^2+2ABz^3)}{32(1-u)^2}+\right.\\
&\left. +\frac{B^3z^3}{16(1-u)^3}-\frac{5uB^3z^3}{48(1-u)^3}+\frac{5u^2B^3z^3}{112(1-u)^3}\right \}.
\end{aligned}
\end{equation}   
To determine $(A,B)$ we compare the original Eq. (\ref{eq:tide}) with Eq. (\ref{TDGE3}) both evaluated for $u=0$ and $R=1$; we find\footnote{For Eq. (\ref{eq:tide}) we used the following limit: $\lim_{x\to +\infty}(x^2-a)^{\frac{1}{2}}-x=0$.}  
\begin{equation} \label{AF} 
1-\cos\alpha=\frac{B}{2}z+\frac{1}{2}\left(A+\frac{B^2}{4}\right)z^2+\frac{B}{4}\left(A+\frac{B^2}{4}\right)z^3,
\end{equation}
where on the left and right hand sides are the results of Eq. (\ref{eq:tide}) and Eq. (\ref{TDGE3}), respectively. By imposing $A+B^2/4=0$ the coefficients of the second and third order in $z$ vanish. Like in the light bending case, Eq. (\ref{AF}) reduces to $1-\cos\alpha=Bz/2$; defining again $B=2$ (implying  $A=-1$) substituting in Eq. (\ref{TDGE3}), we derive the approximate travel time delay equation \citep[see for further details][]{Poutanen06}
\begin{equation} \label{FATIDE}
\frac{\Delta t}{R}=y\left [1+\frac{uy}{8}+\frac{uy^2}{24}-\frac{u^2y^2}{112} \right ],
\end{equation}
where $y=(1-\cos\psi)$.

\begin{figure}[h]
\centerline{\psfig{figure=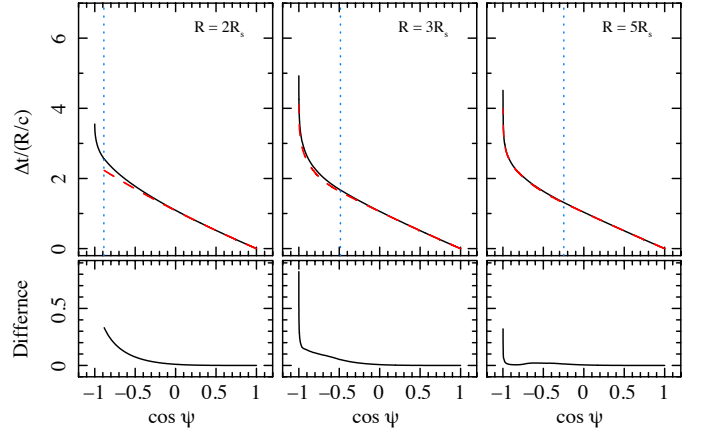,trim=-1cm 3cm 3cm 3.52cm, scale=0.38}} 
\caption{Continuous black curves are obtained from the original time delay Eq. (\ref{eq:tide}), while the dashed red curves are from the polynomial approximate Eq. (\ref{FATIDE}). Different panels show $R=2r_s$, $R=3r_s$ and $R=5r_s$. The dotted blue line helps distinguishing trajectories without a turning point (i.e., $0\le\cos\alpha\le1$) from those with a turning point (i.e., $\pi/2\le\alpha\le\alpha_{max}$). The lower panels show the difference between the curves from the original and approximate equations.} 
\label{fig:Fig3}
\end{figure}

In Fig. \ref{fig:Fig3} we compare for different emission radii the exact travel time delay curves with the polynomial approximated equations. We here also extend the validity of the approximation to $\alpha_{max}$-values accounting for turning points. 
The accuracy settles $\sim35\%$ for $R=2r_s$, while after $R=3r_s$ it is lower than 20\%, according to the same symmetry argument explained in the Sect. \ref{sec:applibe}. However, we refer to \citet{Poutanen06} for the error estimation between $0\le \alpha \le \pi/2$.

\subsection{Solid angle}
\label{sec:appsoan}
We now apply the same method to derive for the first time a polynomial approximation to the solid angle Eq. (\ref{EFSA}). We note, at variance of light bending and time delay equations, that the solid angle equation has the integral in the denominator, and moreover, the emission angle, $\alpha$, is also outside the integral. We first rewrite Eq. (\ref{EFSA}) as 
\begin{equation} \label{INI}
d\Omega=\frac{P_{\rm 1}\ P_{\rm 2}}{I}\ dr\ d\varphi,
\end{equation}
where
\begin{equation} \label{eq:soan}
\begin{aligned}
P_{\rm 1}&=\frac{\cos i}{D^2\sin^2\psi\ (1-u)},\qquad P_{\rm 2}=\frac{\sin^2\alpha}{\cos\alpha},\\
& \\
& I=\int_R^\infty \frac{dr}{r^2}\left[1-\frac{R^2\sin^2\alpha}{r^2(1-u)}\left(1-\frac{uR}{r} \right) \right]^{-\frac{3}{2}}.
\end{aligned}
\end{equation}
$P_{\rm 1}$ is a constant because $\psi$ is a function of the azimuthal angle, $\varphi$, the inclination angle, $i$, and the polar coordinate, $\theta$, (for further details see Sect. \ref{sec:applications}). As a first step, we expand the integrand of $I$ in a Taylor series up to the third order in $z$. We derive
\begin{equation}\label{AI} 
I\approx \frac{1+Cz+Dz^2}{R},
\end{equation}
with
\begin{eqnarray}
C&=&\frac{B}{2(1-u)}-\frac{3Bu}{8(1-u)},\\
D&=&\frac{A}{2(1-u)}-\frac{3Au}{8(1-u)}+\frac{3B^2}{8(1-u)^2}+\\
&+&\frac{15B^2u^2}{16(1-u)^2}-\frac{5B^2u}{8(1-u)^2}. \notag
\end{eqnarray}
The function $P_{\rm 2}/I$ is not yet a polynomial function since it contains a ratio of polynomials and square root functions in $P_{\rm 2}$. For these reasons we expand $P_{\rm 2}/I$ in a Taylor series around $z=0$ and neglect all the terms up to third order in $z$
\begin{equation}\label{AIA} 
\begin{aligned}
\frac{P_{\rm 2}}{I}&\approx \frac{Az^2+Bz}{\sqrt{1-Az^2-Bz}}\frac{R}{1+Cz+Dz^2}\approx\\
&\approx R\left [Bz+\left(\frac{B^2}{2}+A-CB \right)z^2+\right.\\
&\left. +\left(AB+\frac{3B^2}{8}-\frac{CB^2}{2}-CA+BC^2-BD \right)z^3 \right ].
\end{aligned}
\end{equation}
To determine $(A,B)$ we compare the original solid angle Eq. (\ref{eq:soan}) with the above
approximate equation, evaluating both equations for $u=0$ and $R=1$; we find
\begin{equation}
\sin^2 \alpha=Bz+Az^2.
\end{equation}
The left- and right-hand sides are the result of original Eq. (\ref{eq:soan}) and the polynomial Eq. (\ref{AIA}), respectively. We can freely define the value of $A$ and $B$ because there are no particular constraints to impose. We set, as in the previous cases, $A=-1$ and $B=2$, deriving again $z=1-\cos\alpha$. The final approximate equation for the solid angle is
\begin{equation} \label{AFSA} 
\begin{aligned}
d\Omega&\approx \frac{\cos i}{D^2\sin^2\psi\ (1-u)}\ R \left[ 2z+\left(1-2C\right)z^2+\right. \\
&\left. +\left(1-C+2C^2-2D\right)z^3\right]\ dr\ d\varphi,
\end{aligned}
\end{equation}
where
\begin{equation}
C=\frac{4-3u}{4(1-u)},\qquad D=\frac{39u^2-91u+56}{56(1-u)^2}.
\end{equation}

\begin{figure}[ht]
\hbox{
\psfig{figure=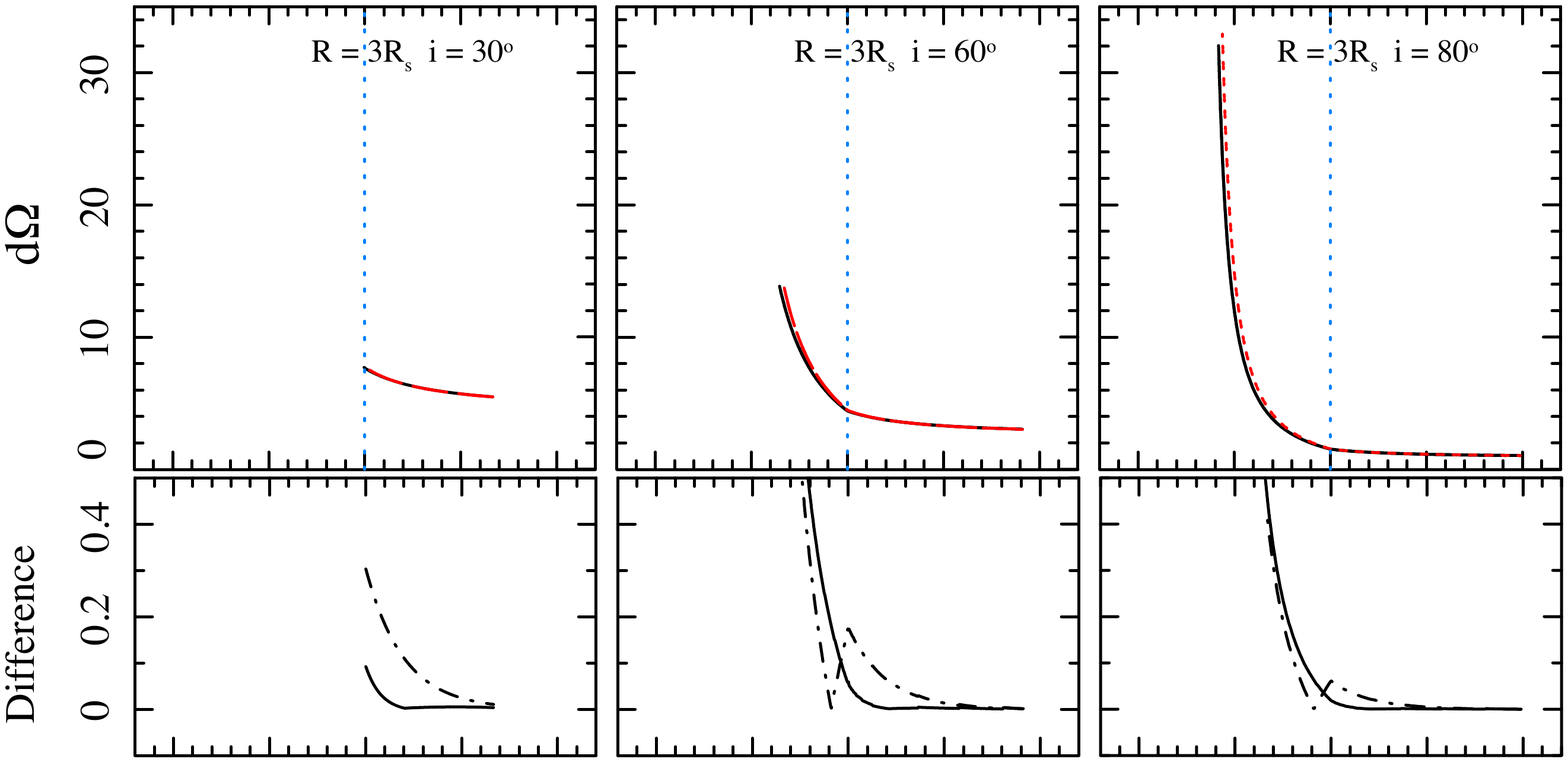, trim=0cm 0cm 4.6cm 1cm, scale=0.326}
}
\vspace{0.05cm}
\hbox{
\psfig{figure=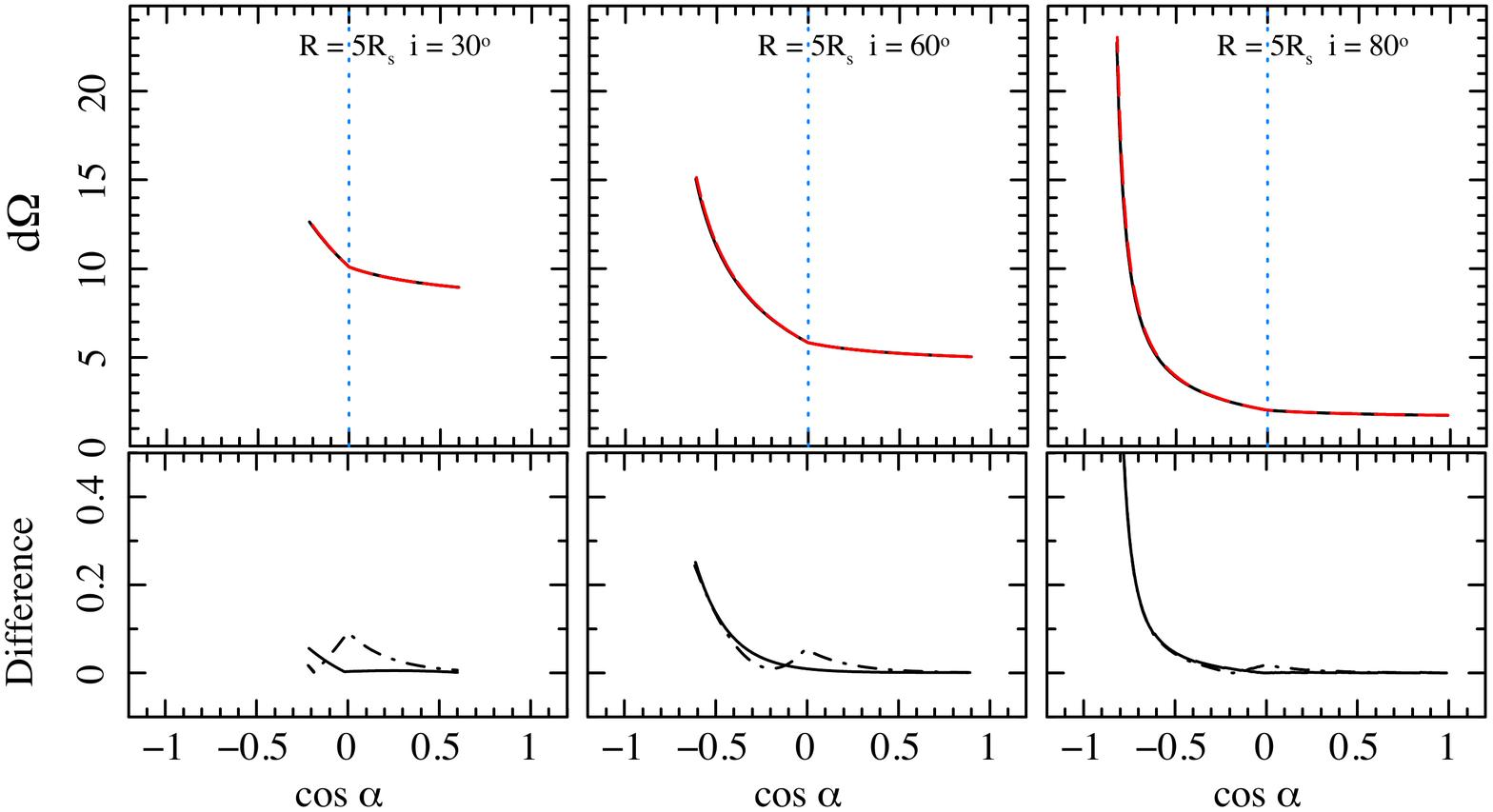, trim=0.3cm 5.5cm 4.6cm 7.97cm, scale=0.326}
}
\vspace{-0.1cm}
\caption{Continuous black curves are obtained from the original solid angle Eq. (\ref{EFSA}), while the dashed red curves are from the polynomial approximation in Eq. (\ref{AFSA}). Different panels are for $R=3r_s$ and $R=5r_s$ and three different inclination angles, $i=30\degr$, $i=60\degr$, and $i=80\degr$. The dotted blue line helps distinguishing trajectories without a turning point (i.e., $0\le\cos\alpha\le1$) from those with a turning point (i.e., $\pi/2\le\alpha\le\alpha_{max}$). The lower panels show the difference between the curves from the original and approximate equations. The dotted-dashed lines represent the difference between the curves from the original and \citet{Beloborodov02} equations.}
\label{fig:Fig4}
\end{figure}

As for the previous two cases, in Fig. \ref{fig:Fig4} we compare the exact solid angle curves with the polynomial approximated curves for different radii and inclination angles $i$. The comparison extends to $\alpha_{max}$-values and thus accounts for trajectories with turning points in this case as well. For $R=3r_s$ the error is $\sim5\%$ and after $R=5r_s$ it is lower than $1\%$. We note that for $i=30\degr$ the curves are fairly flat because the relativistic effects are small. Instead, passing from $i=60\degr$ to $i=80\degr$ , the curves become gradually steeper as general relativistic effects increase. Unlike the previous cases, we do not show here the case $R=2r_s$ because the approximate formula Eq. (\ref{AFSA}) does not give adequately accurate results.

We note that Eq. (A3) in \citet{Beloborodov02} is obtained by approximating the derivative $\frac{d\cos\psi}{d\cos\alpha}$ with the linear Eq. (\ref{AFLB}), while our Eq. (\ref{AFSA}) is a third-order polynomial that approximates the integral $I$ and all the terms depending on the emission angle $\alpha$. For example, our approximation is more accurate by a factor of $\sim$3 to 10 for $R=3r_s$ and $0\le\cos\alpha\le0.3$.
 
\section{Examples of astrophysical applications}
\label{sec:applications}
In this section we present three simple examples of astrophysical applications of the approximate equations. 
We consider the emission point at coordinates $(r,\varphi,\theta)$. The observer is located at infinity along the $z'$-axis with a viewing angle, $i$, with respect to the $z$-axis; the observer polar coordinates are ($r',\varphi',\theta'$). Photons emitted from a point are deflected by an angle, $\psi$, and reach the observer with impact parameter, $b$. The plane containing the photon trajectory rotates around the line of sight as the emission point moves around the compact object. Two unit vectors are attached to the photon emission point, $E$: ${\bf u}$ is tangential to the photon trajectory, and ${\bf n}$ points in the same direction as the radius, $R$. The photon deflection angle, $\psi$, varies as
\begin{equation}
\cos\psi=\sin i\sin\theta \cos\varphi+\cos i \cos\theta,
\end{equation}
with $\theta=\pi/2$, $\varphi = \omega_{k} t$ and $t=0$ when the emission point is closest to the observer. The photon arrival time, $T_{\rm obs}$, is the sum of the emission time, $T_{\rm orb}=\varphi/\omega_{k}$, plus the photon propagation delay, $\Delta T(b)$, from the emission point to the observer (see Eq. (\ref{eq:tide}) ). 

The observed flux is $F=\int_{\nu_{obs}}\int_\Omega I_{\nu_{obs}}d\Omega\ d\nu_{obs}$, where $I_{\nu_{obs}}$ is the specific intensity at the photon frequency $\nu_{obs}$. We use the Lorentz invariant ratio $I_{\nu_{obs}}/\nu^3_{obs}=I_{\nu_{em}}/\nu^3_{em}$ \citep[see, e.g.,][]{Misner73}, where $I_{\nu_{em}}(\xi)=\frac{\epsilon_0 \xi^{q}}{4\pi}\delta(\nu_{em}-\nu_{obs})$ is the specific intensity at the emission point $E$ given by the product of the surface emissivity, varying as a power law of $\xi=R/M$ with index $q$, and the delta function peaked at $\nu_{em}$. Therefore, integrating over all the frequencies, we obtain the observed flux at frequency $\nu_{em}$, $F_{\nu_{em}}=\int_{\Omega} \frac{\epsilon_0 \xi^{q}}{4\pi}\,(1+z)^{-4}\ d\Omega$. The redshift is defined as the ratio between the observed and the emitted energy, $(1+z)^{-1}=\nu_{obs}/\nu_{em}$ \citep{Misner73} and for matter orbiting in circular orbits around a compact object or for a spot on a NS surface reads as
\begin{equation}\label{Redshift2} 
(1+z)^{-1}=\left(1-\frac{r_{\rm s}}{R}-\omega^{2}R^{2}\sin^{2}\theta\right)^{1/2}\left(1+b\omega         
\frac{\sin i\,\sin\varphi\sin\theta}{\sin\psi} \right)^{-1}. 
\end{equation}
For $\omega = \omega_k$ we consider matter orbiting with Keplerian velocity around a BH, and for $\omega = \omega_{spin}$ we consider spots rotating with the NS spin frequency. The relevant geometry is shown in Fig. (\ref{fig:Fig5}).  

\subsection{Light curve from an emitting clump orbiting a black hole}
\label{sec:spotsBH}

\begin{figure*}[ht]
\hbox{
\psfig{figure=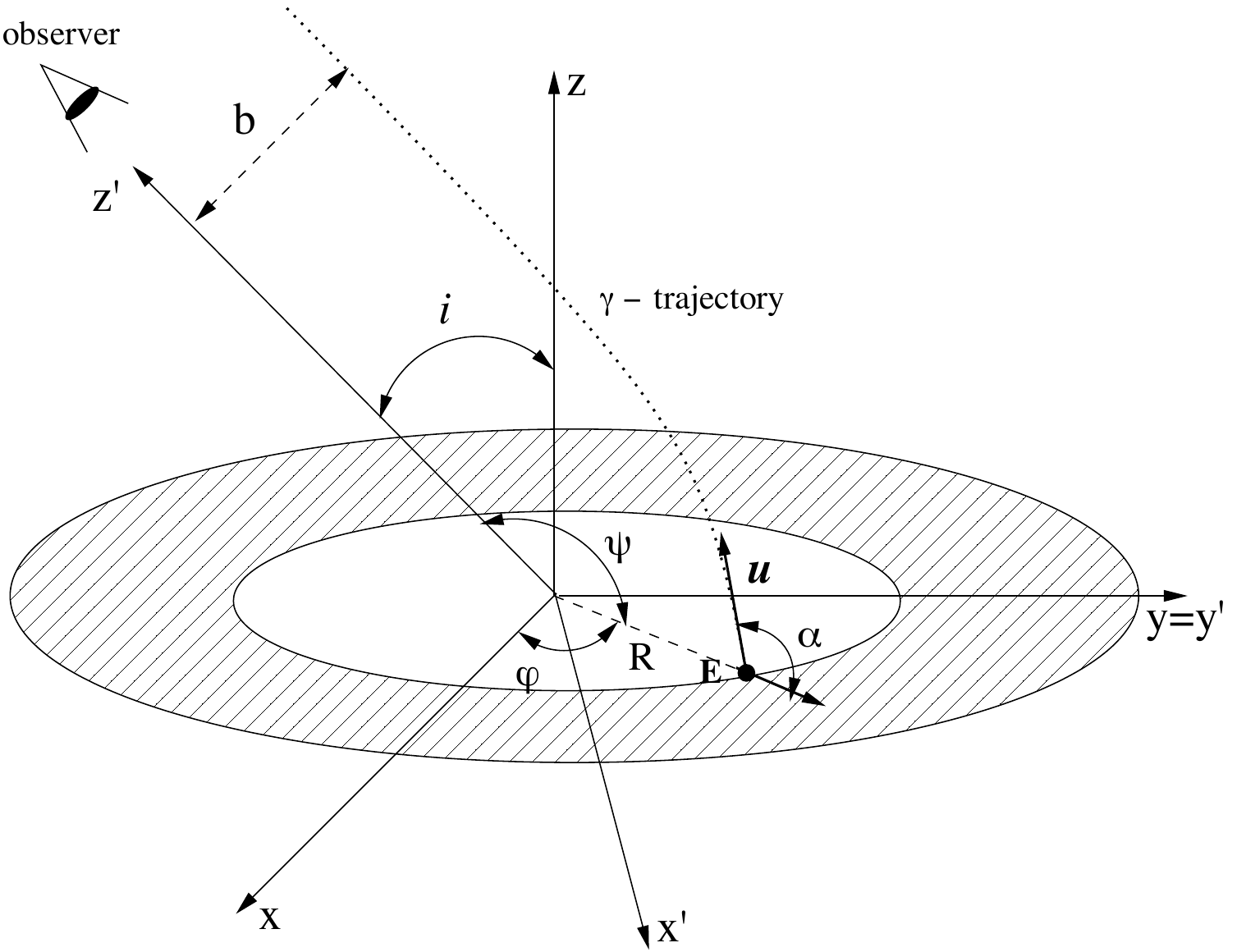, trim=-3.0cm 0cm 4.6cm 2cm, width=6.0cm}
\hspace{5.0cm}
\psfig{figure=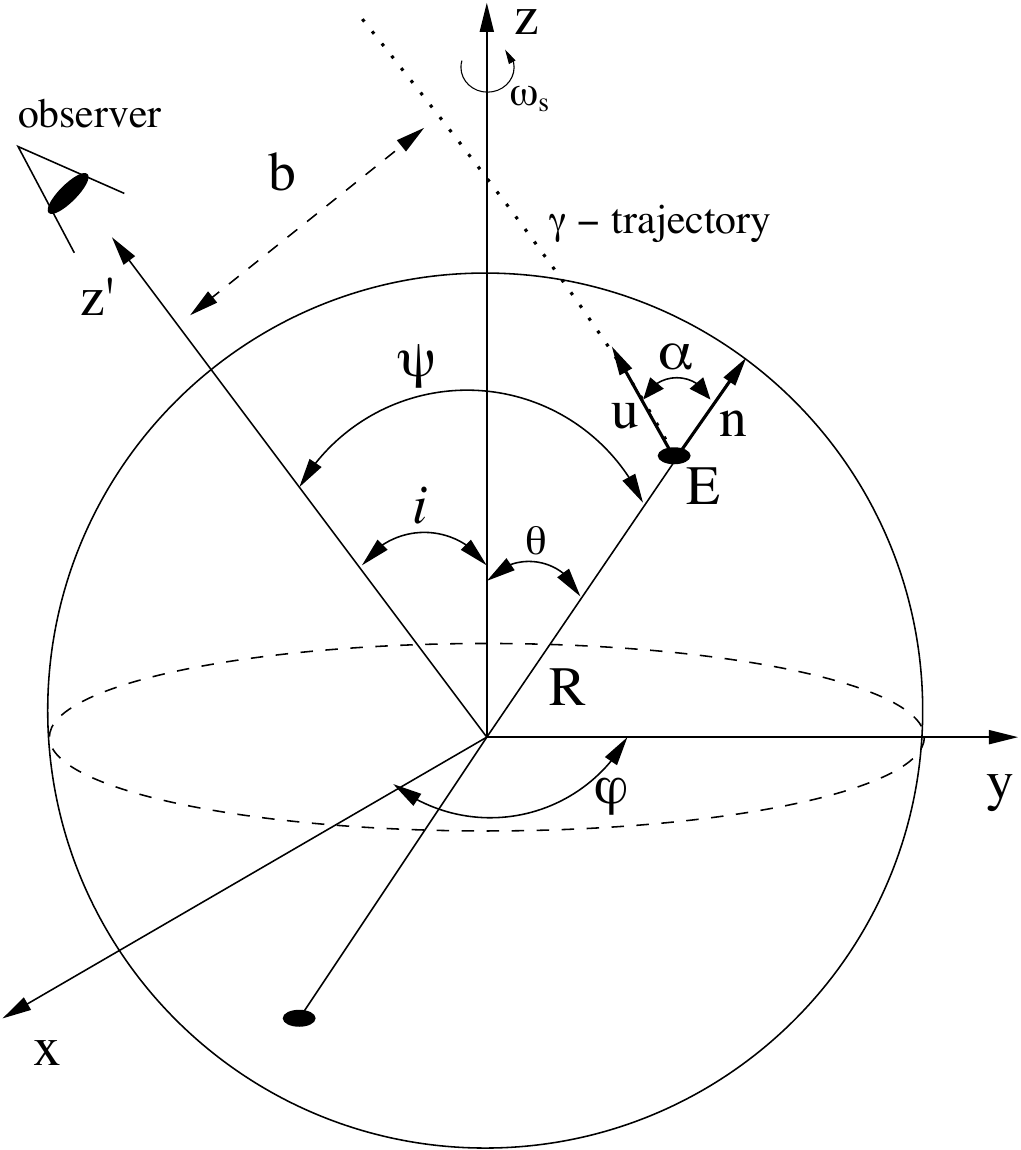, trim=-1cm -0cm 3.6cm 1cm, width=3.5cm}
}
\vspace{0.1cm}
\caption{Geometries adopted in the examples. \emph{Left:}  Emission from a disk or clump orbiting a Schwarzschild BH. \emph{Right:} Emission from two hot opposite spots on an NS surface.} 
\label{fig:Fig5} 
\end{figure*}

\begin{figure}[ht]
\hbox{
\psfig{figure=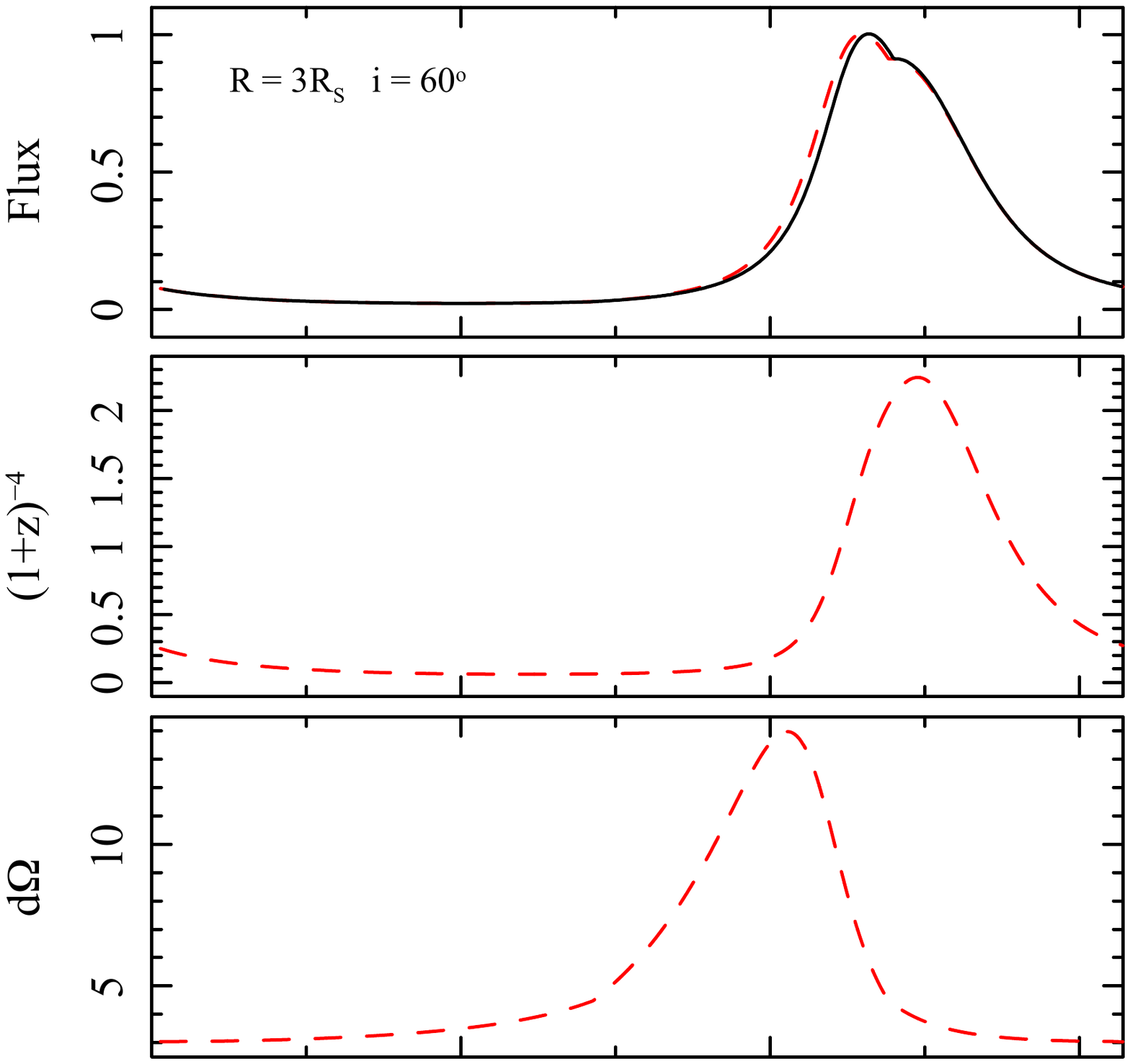, trim=4cm 2cm 4.6cm 1cm, width=5.0cm}
\hspace{-1.4cm}
\psfig{figure=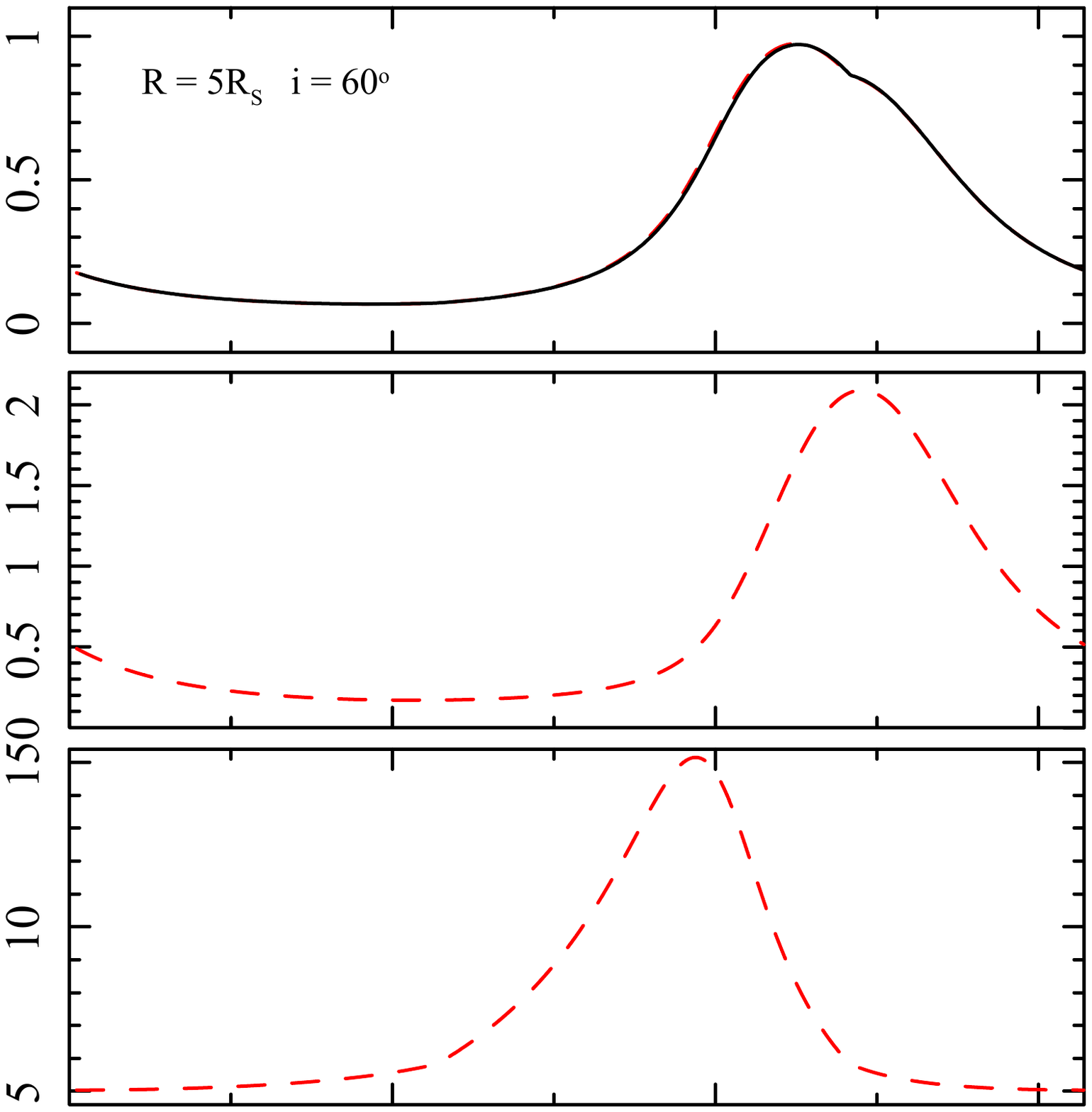, trim=2cm 2cm 6.5cm 1cm, width=5.0cm}
}
\vspace{-0.1cm}
\hbox{
\psfig{figure=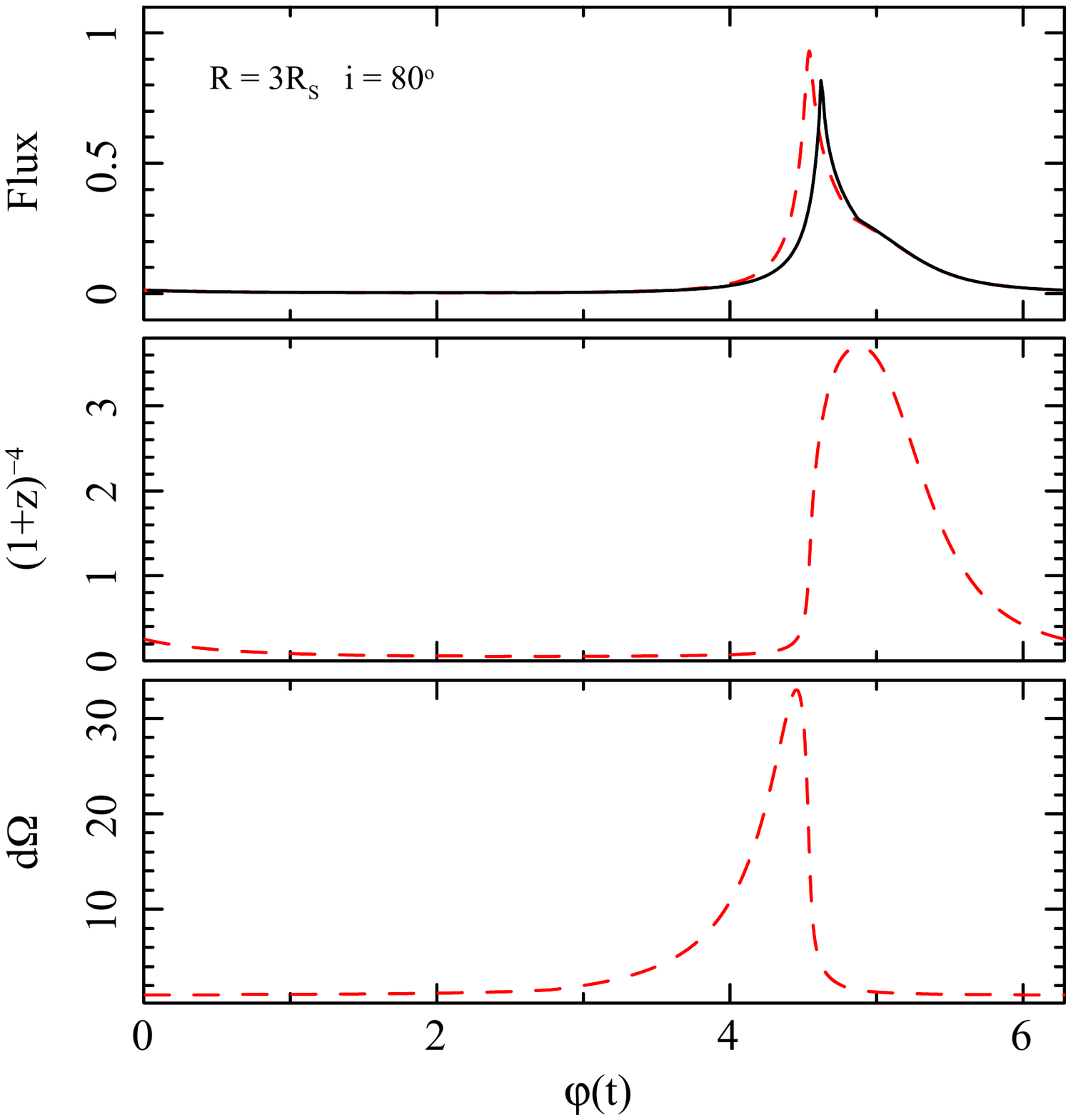, trim=4cm 2cm 4.6cm 2cm, width=5.0cm}
\hspace{-1.4cm}
\psfig{figure=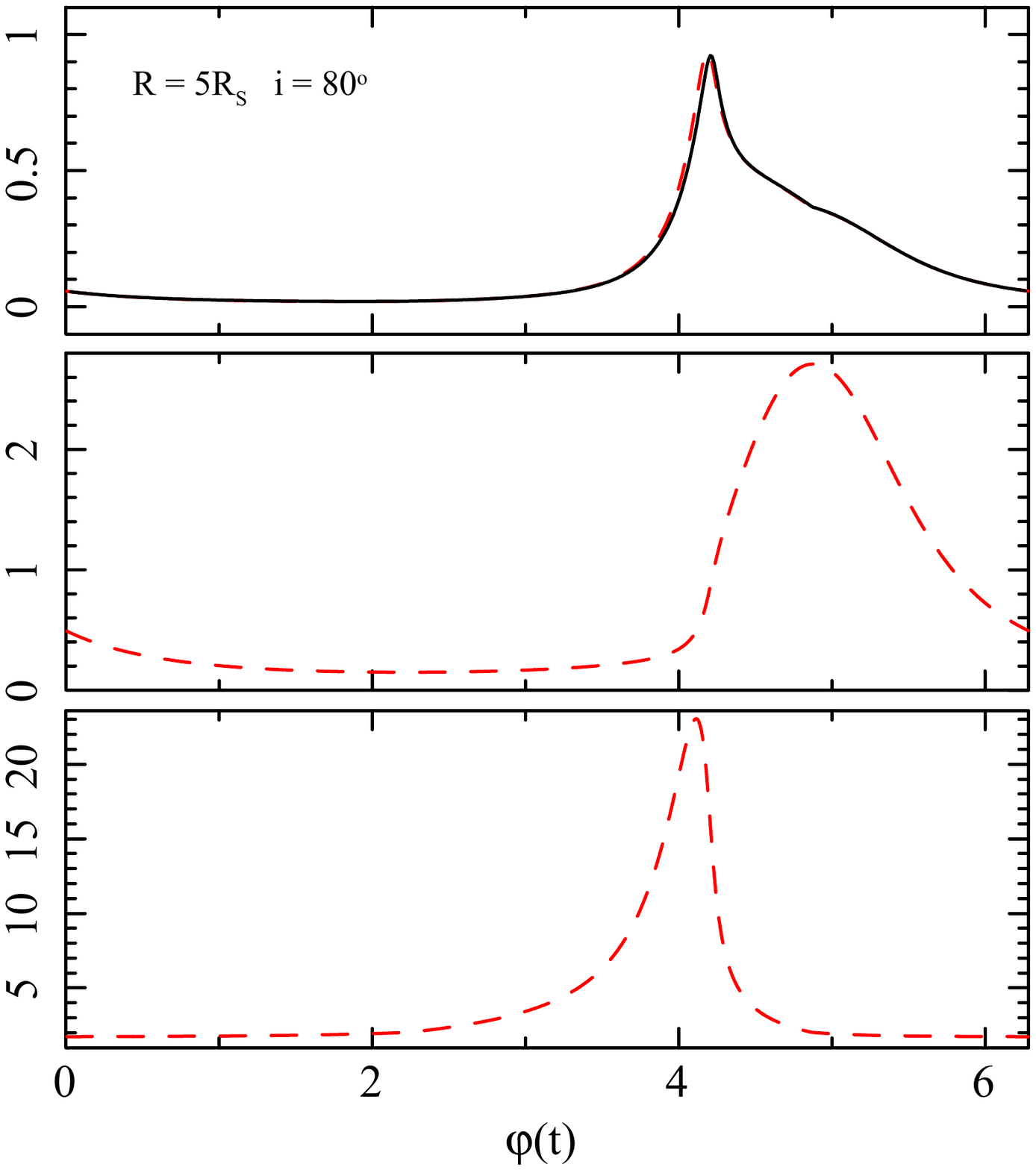, trim=2cm 2cm 6.5cm 2cm, width=5.0cm}
}
\vspace{-0.1cm}
\caption{Modulated flux (normalized to the maximum), Doppler factor $(1+z)^{-4}$ and solid angle (arbitrary units) in the rest coordinate frame of an emitting clump in a circular orbit around a Schwarzschild BH for different radii and inclinations angles. The continuous black lines are calculated with the exact equations, while the dashed red lines are calculated with the approximate equations. All quantities are plotted as a function of the arrival phase at the observer. In the left panels a self-eclipse of the spot is apparent.
} 
\label{fig:Fig6}
\end{figure}

We first consider a clump defined as a small sphere radiating isotropically in its own rest frame, orbiting a Schwarzschild BH in a circular orbit with angular velocity $\omega_{k} = (M/R^{3})^{1/2}$. The geometry is shown in Fig. (\ref{fig:Fig5}). For simplicity we assume $\frac{\epsilon_0 \xi^{q}}{4\pi}=1$. 
Figure (\ref{fig:Fig6}) shows the modulation of the Doppler factor $(1+z)^{-4}$, solid angle $d\Omega,$ and flux from the orbiting clump as a function of phase, $\varphi(t)$, including light travel time delays. When the clump is behind the BH, gravitational lensing magnifies the solid angle from which the clump is seen by observer; the Doppler factor is greatest when the projected velocity along the photon trajectory reaching the observer is highest. The gravitational effects are stronger for larger inclination angles, and the observed peak flux is not at $\varphi=\pi$, but is significantly shifted especially for large inclination angles due to the travel time delays. The errors between the approximated and the original flux depend only on the emission radius, since the inclination angle figures as a constant. However, it is evident that the main errors derive from the approximated time delay equation (as shown in the Sect. \ref{sec:apptide}). 

\subsection{Emission line profile from an accretion disk around a black hole}
\label{sec:Ironline}
\begin{figure}[h]
\centerline{\psfig{figure=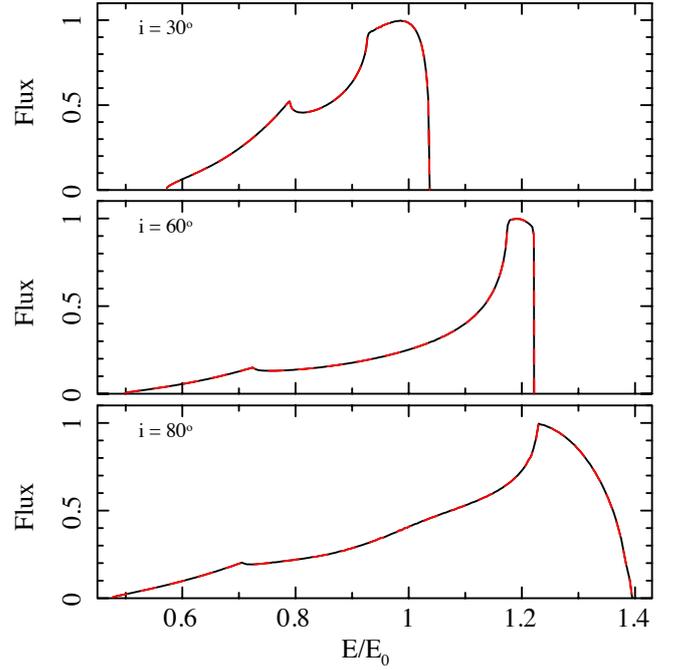,trim=2cm 2cm 6.5cm 2cm,width=9.5cm}} 
\caption{Line profile for isotropic radiation from $R_{\rm in}=3\,r_{\rm s}$ to $R_{\rm out}=50\,r_{\rm s}$ assuming surface emissivity $q=-3$. The continuous black lines represent the original equations and the dashed red lines are the polynomial approximate equations.} 
\label{fig:Fig7}
\end{figure}

In Fig. (\ref{fig:Fig7}) we calculate the steady relativistically broadened emission line profile from an accretion disk around a Schwarzschild BH \citep[e.g.,][and references therein]{Fabian89,beckwith04}. Fe $K{\rm \alpha}$ lines at $\sim 6- 7$~keV from a number of accreting stellar mass BHs and NSs in X-ray binaries, as well as supermassive BHs in the nuclei of active galaxies are interpreted based on this model \citep[e.g.,][]{tomsick14}. We integrate over the disk surface from an inner to an outer disk radius and ignore light propagation delays, as we consider a steady disk. The approximate equations reproduce very accurately the profiles obtained with the exact equations.  A high accuracy is also retained for large inclination angles, even if larger inclination angles enhance the relativistic effects (see Sect. \ref{sec:spotsBH}). 
  
\subsection{Light curve from a hot spot on the surface of a rotating neutron star}
\label{sec:spotsNS}
\begin{figure}[ht]
\hbox{
\psfig{figure=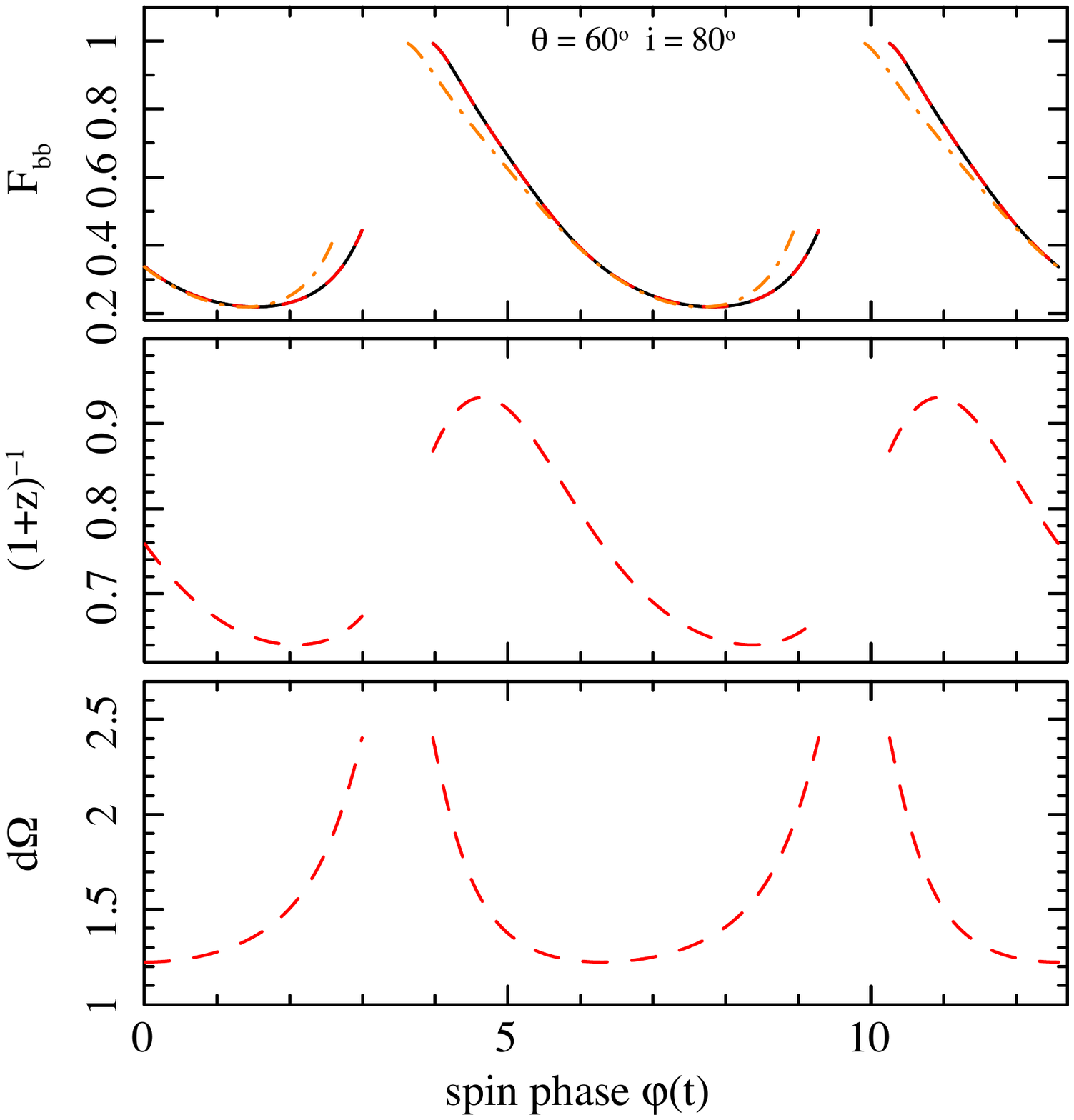, trim=4cm 2cm 4.6cm 1cm, width=5.2cm}
\hspace{-1.4 cm}
\psfig{figure=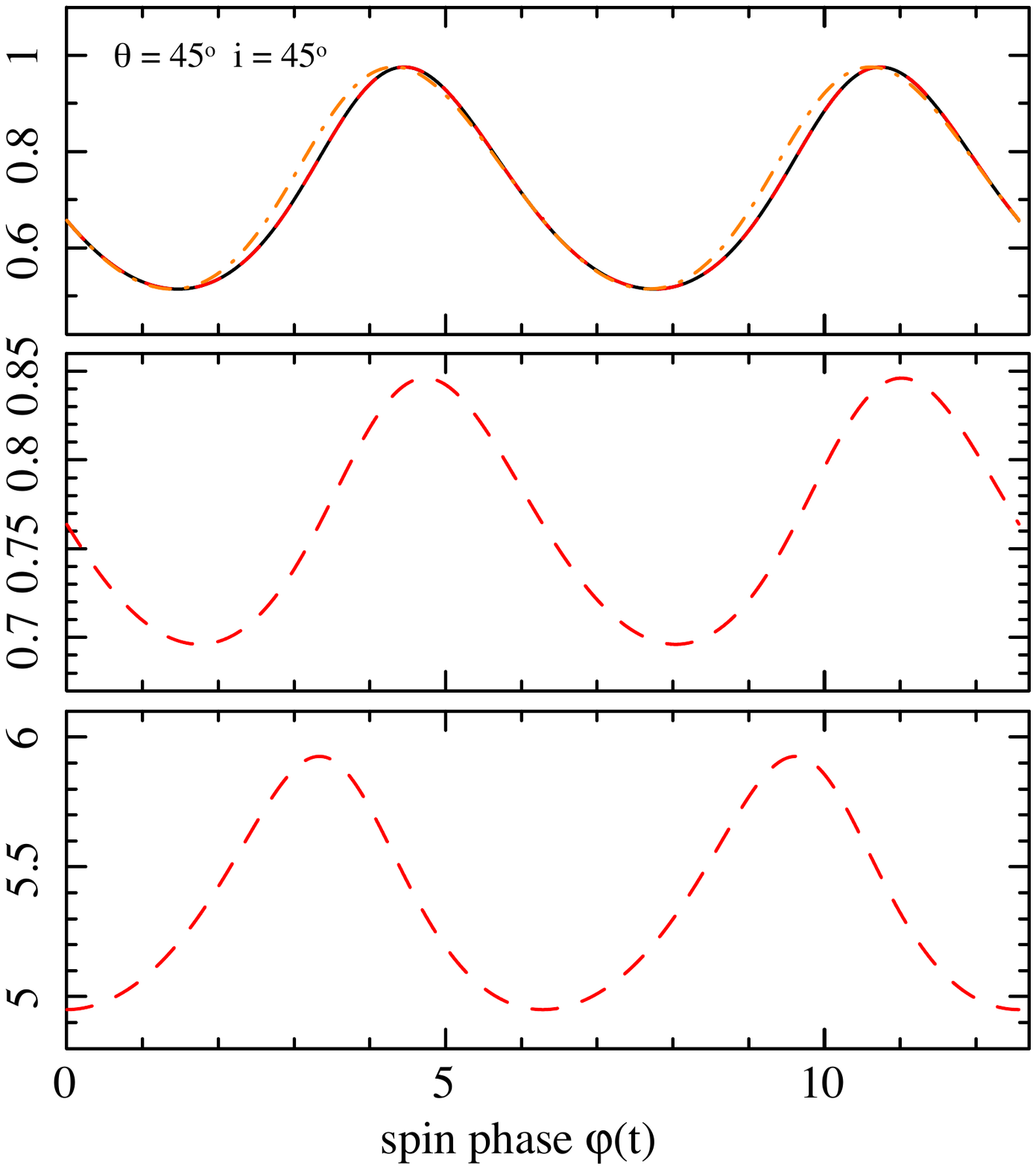, trim=2cm 2cm 6.5cm 1cm, width=5.2cm}
}
\caption{Modulation from a hot spot on an NS as a function of rotational phase for different inclination angles and hot spot colatitude. Light travel time delays are included.  The continuous black lines represent the results from a numerical integration of the original equations; the dashed red lines are obtained from the polynomial approximate equations. The dashed-dotted orange line does not include light travel time delays.} 
\label{fig:Fig9}
\end{figure}

We calculate here the pulse profile generated by a point-like hot spot located on the surface of a NS, which emits like an isotropic blackbody.  Calculations of this type have been carried out extensively to 
model the periodic signals of accreting millisecond pulsars \citep[see, e.g.,][] {Pechenick83,Poutanen06,leahy11,baubock15} as well as the so-called burst oscillations during Type I thermonuclear bursts in NS low-mass X-ray binaries \citep[e.g.,][]{nath02,miller15}; some of these  calculations also include the angular size of the hot spot, the star oblateness, and the spacetime modifications induced by fast rotation. We use here a canonical NS mass of $1.4M_{\odot}$ and radius $R_{\rm NS} = 12$ km, together different inclination angles, $i$, and colatitudes, $\theta$ of the spot. The NS spin frequency is chosen to be $\nu_s = 600$ Hz. In Fig. (\ref{fig:Fig9}) we report the corresponding pulse profiles; as expected, the case with higher values
of $i$ and $\theta$ displays larger departures from a sinusoidal shape. In this type of applications the value of  $\alpha$ is always limited to $\leq \pi/ 2$, as no turning points are involved. Therefore our approximate equations retain very high accuracy as long as the NS radius is $ \ge 2.5r_s$, a range that encompasses a number of NS 
models for different equations of state, excluding only the upper end of the mass-radius branches. We conclude that our approximate equations can be usefully employed in calculations of the pulse profile of fast spinning NSs over a range of (but not all) models to be tested against the observation that the Neutron Star Interior Composition ExploreR (NICER), and other large-area X-ray missions of the future, such as Athena or LOFT, will obtain \citep[see][and references therein]{watts16}.

\subsection{Applicability regions}
\label{sec:appreg}
\begin{figure}[ht]
\centerline{\psfig{figure=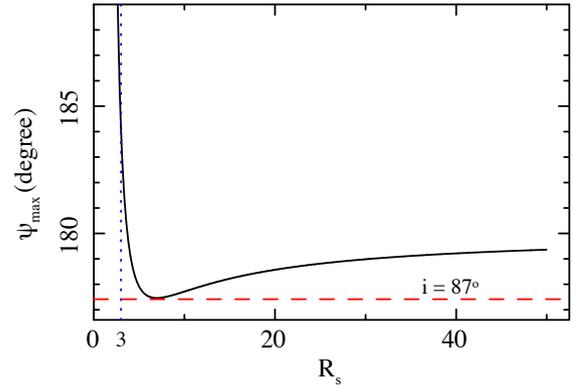,trim=-2cm 2cm 6.5cm 8cm,width=11cm}} 
\caption{Largest bending angle $\psi_{max}$, vs. the emission radius (continuous black line). For inclination angles below $i=87\degr$ (dashed red line) the approximate equations provide a high accuracy, since they are below the $\psi_{max}$-value. $R=3r_s$ (dotted blue line) separates the applicability region from the strong-field regime ($R<3r_s$).} 
\label{fig:Fig10}
\end{figure}

In Fig. (\ref{fig:Fig10}) we plot $\psi_{\rm max}$ as a function of the emission radius to investigate the applicability regions of the approximate equations. If we consider trajectories with turning points for radii $R<3r_s$, that is,{\it } smaller than the ISCO, then $\psi_{\rm max} \ge 180\degr$ and a polynomial treatment is no longer accurate because of strong field effects (see also Fig. (\ref{fig:Fig2}) ). We note that for $R\longrightarrow1.5r_s$, $\psi_{\rm max}$ our solution approaches asymptotically $270\degr$. Instead, for $R\ge3r_s$, when the observer is located edge on (i.e., $i=90\degr$), $\psi_{\rm max}=180\degr$ is attained; otherwise, for slightly smaller but still extreme inclination angles, for
example,{\it } $87\degr$, photon trajectories always remain below the critical bending angle, which guarantees a high accuracy of our polynomial approximations. This argument is valid for all the emission radii $R\ge3r_s$, since for $R \longrightarrow  \infty$, $\psi_{\rm max}$ approaches $180\degr$. 

\section{Conclusions}
\label{sec:conclusions}
We developed an analytical method to approximate the elliptic integrals that describe gravitational light bending and light travel time delays of photon geodesics in the Schwarzschild metric. Based on this, we derived for the first time an approximate polynomial equation also for the solid angle. We discussed the accuracy and range of applicability of the approximate Eqs. (\ref{AFLB}), (\ref{FATIDE}), and (\ref{AFSA}); adopting them can considerably speed up calculations related to a variety astrophysical problems, which normally require time-consuming numerical integrations. We also presented a few simple applications as examples. We will extend our treatment to the parallel transport of polarization vectors in a future work. 

\section*{Acknowledgements}
This research was financed by the Swiss National Science Foundation project 200021\_149865. VdF and MF acknowledge the Department of Physics at the University of Basel and especially Friedrich-K. Thielemann. We also thank the International Space Science Institute in Bern for their support. VdF is grateful to the International Space Science Institute in Beijing for the hospitality to carry out part of this work. LS acknowledges partial support under contract ASI INAF I/004/11/1.

\bibliographystyle{aa}
\bibliography{references}

\begin{thebibliography}{15}
\expandafter\ifx\csname natexlab\endcsname\relax\def\natexlab#1{#1}\fi

\bibitem[{{Bao} {et~al.}(1994){Bao}, {Hadrava}, \& {Ostgaard}}]{Bao1994}
{Bao}, G., {Hadrava}, P., \& {Ostgaard}, E. 1994, \apj, 435, 55

\bibitem[{{Baub{\"o}ck} {et~al.}(2015){Baub{\"o}ck}, {Psaltis}, \&
  {{\"O}zel}}]{baubock15}
{Baub{\"o}ck}, M., {Psaltis}, D., \& {{\"O}zel}, F. 2015, \apj, 811, 144

\bibitem[{{Beckwith} \& {Done}(2004)}]{beckwith04}
{Beckwith}, K. \& {Done}, C. 2004, \mnras, 352, 353

\bibitem[{{Beloborodov}(2002)}]{Beloborodov02}
{Beloborodov}, A.~M. 2002, \apjl, 566, L85

\bibitem[{{Chandrasekhar}(1992)}]{Chandrasekhar92}
{Chandrasekhar}, S. 1992, {The mathematical theory of black holes}

\bibitem[{{Fabian} {et~al.}(1989){Fabian}, {Rees}, {Stella}, \&
  {White}}]{Fabian89}
{Fabian}, A.~C., {Rees}, M.~J., {Stella}, L., \& {White}, N.~E. 1989, \mnras,
  238, 729

\bibitem[{{Leahy} {et~al.}(2011){Leahy}, {Morsink}, \& {Chou}}]{leahy11}
{Leahy}, D.~A., {Morsink}, S.~M., \& {Chou}, Y. 2011, \apj, 742, 17

\bibitem[{{Luminet}(1979)}]{Luminet79}
{Luminet}, J.-P. 1979, \aap, 75, 228

\bibitem[{{Miller} \& {Lamb}(2015)}]{miller15}
{Miller}, M.~C. \& {Lamb}, F.~K. 2015, \apj, 808, 31

\bibitem[{{Misner} {et~al.}(1973){Misner}, {Thorne}, \& {Wheeler}}]{Misner73}
{Misner}, C.~W., {Thorne}, K.~S., \& {Wheeler}, J.~A. 1973, {Gravitation}

\bibitem[{{Nath} {et~al.}(2002){Nath}, {Strohmayer}, \& {Swank}}]{nath02}
{Nath}, N.~R., {Strohmayer}, T.~E., \& {Swank}, J.~H. 2002, \apj, 564, 353

\bibitem[{{Pechenick} {et~al.}(1983){Pechenick}, {Ftaclas}, \&
  {Cohen}}]{Pechenick83}
{Pechenick}, K.~R., {Ftaclas}, C., \& {Cohen}, J.~M. 1983, \apj, 274, 846

\bibitem[{{Poutanen} \& {Beloborodov}(2006)}]{Poutanen06}
{Poutanen}, J. \& {Beloborodov}, A.~M. 2006, \mnras, 373, 836

\bibitem[{{Tomsick} {et~al.}(2014){Tomsick}, {Nowak}, {Parker}, {Miller},
  {Fabian}, {Harrison}, {Bachetti}, {Barret}, {Boggs}, {Christensen}, {Craig},
  {Forster}, {F{\"u}rst}, {Grefenstette}, {Hailey}, {King}, {Madsen},
  {Natalucci}, {Pottschmidt}, {Ross}, {Stern}, {Walton}, {Wilms}, \&
  {Zhang}}]{tomsick14}
{Tomsick}, J.~A., {Nowak}, M.~A., {Parker}, M., {et~al.} 2014, \apj, 780, 78

\bibitem[{{Watts} {et~al.}(2016){Watts}, {Andersson}, {Chakrabarty}, {Feroci},
  {Hebeler}, {Israel}, {Lamb}, {Miller}, {Morsink}, {{\"O}zel}, {Patruno},
  {Poutanen}, {Psaltis}, {Schwenk}, {Steiner}, {Stella}, {Tolos}, \& {van der
  Klis}}]{watts16}
{Watts}, A.~L., {Andersson}, N., {Chakrabarty}, D., {et~al.} 2016, Reviews of
  Modern Physics, 88, 021001

\end{thebibliography}

\end{document}